\documentstyle[11pt,epsf,amsfonts,amssymb,amsbsy,colordvi]{article}
\thicklines

\newcommand\VV{\setbox0=\hbox{V}\hbox{\rm V\raise\ht0
  \hbox to0pt{\hss\vbox to0pt{\hbox{v}\vss}}}}
\def\slashchar#1{\setbox0=\hbox{$#1$}           
   \dimen0=\wd0                                 
   \setbox1=\hbox{/} \dimen1=\wd1               
   \ifdim\dimen0>\dimen1                        
      \rlap{\hbox to \dimen0{\hfil/\hfil}}      
      #1                                        
   \else                                        
      \rlap{\hbox to \dimen1{\hfil$#1$\hfil}}   
      /                                         
   \fi}                                         %

\begin{document}

\centerline{\uppercase{\large\bf CP-violation in the heavy quark
systems}}

\vspace*{3mm}
\centerline{\sf V.V.Kiselev}
\centerline{\it Institute for High Energy Physics, Protvino, Russia}

\begin{abstract}
We present a review of general picture in the sector of electroweak symmetry
breaking with the CP-violation in the heavy quark interactions.
\end{abstract}

\section*{Introduction}

After the precision measurements of electroweak parameters at LEP, FNAL and
SLAC, the main progress in the experimental study of Standard Model (SM) is
connected to investigations in the Higgs sector. The Higgs mechanism provides
the basic renormalization properties of the SM as well as the masses of vector
gauge fields and fermions. There are two directions in these studies. The first
is an observation of neutral scalar Higgs particle in order to prove the
completeness of SM. The second direction is an investigation of Yukawa
couplings of Higgs scalar with the quark and lepton fields. When the number of
generations is equal to three (as measured to the moment) or greater, these
couplings can be complex, which inavitably leads to the violation of combined
CP parity inverting both the charges (C) and space orientation (P). This
violation after the transition to the observed mass-flavor states of fermions,
manisfests in the mixing of weak charged quark currents. Therefore, measuring
the CKM elements in heavy quark decays tells us about the Higgs sector of SM.

Theoretically, two phenomenological approaches are developed in order to study
the Yukawa sector. The first is a modelling of the mass matrices regardless of
SM extensions. This approach is closely related with the pionering paper by
Fritzsch \cite{Fritzsch1978} devoted to the mass matrix textures. The second
way is based on restricting the Yukawa sector of SM extensions by observed
\underline{regularities}. These regularities of quark current mixing are quite
definite and bright. Indeed, in the nature we deal with 
\begin{itemize}
\item
a single heavy major generation and two almost {massless} junior generations,
and
\item
a small mixing of major generation with the junior generations.
\end{itemize}
These observations are refered to as the hierarchy of masses and hierarchy of
mixings. The hierarchies are combined in the principle of democracy in the
Yukawa interactions of quarks. According to this principle a leading
contribution to the Yukawa interactions involves the only universal coupling
$\lambda_{\rm Ferm}$ for all of three generations composed by equal-charge
fermions, so that the mass matrix has the form, which can be transformed from
the democratic basis to the heavy one in the following way:
$$
M = \lambda_{\rm Ferm}\,\eta_{\rm vac} \left( \begin{array}{ccc} 1 & 1 & 1\\ 1
& 1 & 1\\ 1 & 1 & 1
\end{array}\right) \Longrightarrow M_U = \lambda_{\rm Ferm}\,\eta_{\rm vac}
\left( \begin{array}{ccc} 1 & 0 & 0\\ 0 & 0 & 0\\ 0 & 0 & 0
\end{array}\right),
$$
due to the operation
$$
U\cdot M \cdot U^{\dagger} = M_U,
$$
where the rotation matrix $U$ has the form
$$
U = \left(\begin{array}{ccc} 
1/\sqrt{3} & 1/\sqrt{3} & 1/\sqrt{3} \\
1/\sqrt{2} & -1/\sqrt{2} & 0 \\
1/\sqrt{6} & 1/\sqrt{6} & -2/\sqrt{6} \end{array}\right).
$$
Since the mass matrices for the up-kind and down-kind fermions have the same
form for the unit matrix of charged currents, then after the rotation the
mixing Cabibbo--Kobayashi--Maskawa matrix (CKM) is transformed to the
following:
$$
\hspace*{1.7cm}
V_{CKM} = \left(\begin{array}{ccc}
V_{ud} & V_{us} & V_{ub}  \\
V_{cd} & V_{cs} & V_{cb} \\
V_{td} & V_{ts} & V_{tb} 
\end{array} \right) \Longrightarrow U_{up} U^\dagger_{down} = I,
$$
and in the limit of exact democracy we arrive to the single heavy quark
generation and zero mixing of light junior generations with the major one.

The democracy is the approximate symmetry of Yukawa interactions. In the nature
it is slightly perturbed, that leads to the observed picture of light junoir
generations with small mixing of charged currents. We stress that
\underline{an origin of democracy and its perturbations are
guided by} \underline{a structure beyond the SM.}

Therefore, the study of the heavy quark mixing in weak interactions and the
CP-violation informs us not only about the parameters of SM, but also about the
dynamics underlying the Yukawa interactions beyond the SM.

\section{CKM matrix in SM}
The mixing of charged quark currents in the SM is described by the unitary CKM
matrix with three real rotation angles and a single complex phase providing the
violation of CP invariance. According to the observed hierarchy of mixing, the
elements of this matrix is classified by the order of magnitude. Next step is a
relation of mass-hierarchy regularities with the mixing.

\subsection{Parametrizations \& Textures of quark mass matrices}

Wolfenstein \cite{Wolfenstein} gave a general arrangement of unitary
four-parameter matrix in terms of a small parameter $\lambda = |V_{us}|\approx
0.22$ representing the sine of Cabibbo angle in the mixing of two junior
generations, so that
$$
V_{CKM} = \left(\begin{array}{ccc}
V_{ud} & V_{us} & V_{ub}  \\
V_{cd} & V_{cs} & V_{cb} \\
V_{td} & V_{ts} & V_{tb} 
\end{array} \right)= \left(\begin{array}{ccc}
1-\frac{\lambda^2}{2} & \lambda & A \lambda^3 (\rho- i \eta) \\
-\lambda & 1-\frac{\lambda^2}{2} & A\lambda^2 \\
A \lambda^3 (1-\rho- i \eta) & -A\lambda^2 & 1
\end{array} \right),
$$
where numerically the constant $A$ is about unit, while the $\rho,\, \eta$
parameters are in the range of $0.2-0.3$. This representation of CKM matrix is
completely phenomenological, and it does not involves any assumption on the
nature of parameters.

Fritzsch, Xing \cite{FritzschXing} and Rasin \cite{Rasin} considered a
classification of parametrizations under the limits of small mixings and mass
hierarchy. They found {\bf 9} variants resulted in all of the best form
$$
V_{CKM} = \left(\begin{array}{ccc}
s_u s_d c + c_u c_d e^{-i\tilde\delta} & s_u c_d c - c_u s_d e^{-i\tilde\delta}
& s_u s \\
c_u s_d c - s_u c_d e^{-i\tilde\delta} & c_u c_d c + s_u s_d e^{-i\tilde\delta}
& c_u s \\
-s_d s & -c_d s & c 
\end{array} \right),
$$
where we have used the notations; $s_{u,d} =\sin \theta_{u,d}$, $c =
\cos\theta_3$ and so on. This Fritzsch--Xing parametrization is different from
the standard form given by Kobayashi and Maskawa. The advantage of
Fritzsch--Xing form is due to the transparent relations between the
regularities of quark mass-matrices and the mixing parameters. Indeed,
Fritzscha and Xing shown that, for example, the limit of small mixing of major
generation with the junior ones, i.e. the decoupling of heavy generation, takes
a simplest form for the infinitely heavy major generation. So, 
$$
\sin\theta_3\to 0 \;\;\; \Leftrightarrow \;\;\;  m_{\rm heavy} \to \infty
\;\;\;
\Longrightarrow  \;\;\;\mbox{decoupling of heavy generation.}
$$
The other relation considers the limit of massless junior generation and its
connection to the decoupling and CP-violation, so that 
$$
\sin\theta_{u,d}\to 0 \;\;\; \Leftrightarrow  \;\;\; m_{u,d} \to 0  \;\;\;
\Longrightarrow \;\;\; \left\{\begin{tabular}{l}decoupling of junior
generation,\\ no CP-violation.\end{tabular}\right.
$$
Some other conditions making the preference for the Fritzsch--Xing
parametrization are also discussed in \cite{FritzschXing}. 

\begin{figure}[th]
\centerline{\epsfxsize=5cm \epsfbox{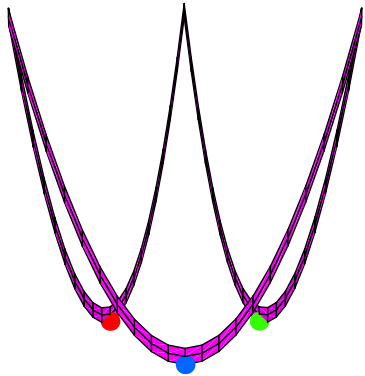}}

\caption{Three equivalent positions in the Higgs sector composing the real
${\mathbb Z}_3$ symmetric vacuum.}
\label{z3}
\end{figure}

The above form of mixing matrix can be easily obtained by the diagonalization
of mass matrices given by \cite{VVK}
\begin{equation}
M = \left(\begin{array}{ccc} 
\mu_1 & \bar\mu+\Delta & \bar\mu \\
\bar\mu+\Delta & \mu_2 & \bar\mu-\Delta \\
\bar\mu & \bar\mu-\Delta & \mu_3 \end{array}\right).
\label{m-cond}
\end{equation}
where the complex parameters are arranged, so that $\mu_{1,2,3} \approx \bar\mu
\gg \Delta$.
Further, one can get the form of (\ref{m-cond}) by a transformation of
quite symmetric original mass matrix \cite{VVK}
\begin{equation}
M_{RL} = \left( \begin{array}{ccc} 
v_1 & v_2 & v_3 \\
v_2 & v_3 & v_1 \\
v_3 & v_1 & v_2  
\end{array}\right).
\label{m3}
\end{equation}
Matrix (\ref{m3}) possesses the permutational symmetry of indices, that does
not change the eigen-values. Its form can be derived from the ${\mathbb Z}_3$
symmetry in the Higgs vacuum sector, which is symbollically shown in Fig.
\ref{z3}.
This symmetry implies that the fermions are coupled to the scalar fields, which
have the same vacuum expectations except the variation of complex phase rotated
by the angle $\frac{2\pi}{3}$. Thus, we deal with\\
$\left.
\begin{array}{r}
\mbox{three real parameters \& one complex phase}\\
{\mathfrak Im}\, v_1 \neq 0,\;\;\; v_{2,3}\in {\mathfrak R}
\end{array}\right\} \Longrightarrow $
\begin{tabular}{l}{\fbox{${\mathbb Z}_3$}
symmetry of vacuum: }\\ $\{a=e^{\frac{2\pi}{3}},\, 1,\,
a^{-1}=e^{-\frac{2\pi}{3}}\}$.
\end{tabular}

It is worth to stress that this picture of vacuum structure is quite general,
and it is can be consistent not only with the quark sector, but also with the
charged leptons and neutrinos. So, if the only admissible complex phase of
$v_1$ is small, i.e. the phase is given by the unit element of ${\mathbb Z}_3$,
then we arrive to the situation with quarks and charged leptons, while in the
case of phase close to the value given by the basis element of ${\mathbb Z}_3$
we get the most probable picture in the neutrino sector with the almost
degenerate neutrinos and large mixings:\\[5mm]
\hspace*{5mm}$|{\mathfrak Im}\, v_1| \ll |v_{1,2,3}|\approx v\;\;\;
\Longrightarrow $
{\fbox{Hierarchy} of charged fermion masses \& mixings.}\\[4mm]
$v_1 = |v_1|e^{\frac{2\pi}{3}},\; |v_{1,2,3}|\approx v\;\;\; \Longrightarrow $
{Almost \fbox{degenerate} neutrinos \& \fbox{large mixing.} } \\

In order to illustrate the framework of symmetric vacuum state, we give the
example with two generations involving the ${\mathbb Z}_2$ symmetry, which
results in two-parametric mass matrix. It can be represented in the symmetric
form transformed to the matrix similar to (\ref{m-cond})
\begin{equation}
M_{RL} =  \left( \begin{array}{cc} 
v_2 & v_1 \\ v_1 & v_2 \end{array}
\right)\; \Longrightarrow \;\; \left( \begin{array}{cc} v-a & v\\ v & v+a
\end{array} \right),
\label{f11}
\end{equation}
where the real parameters $v_{1,2} \in {\mathfrak R}$ are close to each other,
and we have introduced the notations {$v=v_2$, $a=v \tan 2 \theta$}, so that we
get the hierarchy of masses
\begin{equation}
m_{1,2} = |v_1\pm v_2|\;\; \Longrightarrow\;\; m_1 \ll m_2,
\label{f13}
\end{equation}
while mixing angle of generations is related to the mass ratio
$$
\cos 2\theta = \frac{v_2}{v_1},\;\; \Leftrightarrow \;\; 
\tan 2\theta = 2 \frac{\sqrt{m_1 m_2}}{m_2-m_1}\approx 2
\sqrt{\frac{m_1}{m_2}},
$$
that approximately leads to
$$
\displaystyle \sin \theta \approx \sqrt{\frac{m_1}{m_2}}.
$$

\noindent
Similar relations for three generations appear for the mixing of junior
generations, while the mixing with the major one is suppressed as $m_2/m_3$ and
the CP-violation phase is fixed in the case of ${\mathbb Z}_3$. So, we get
$$
\Violet{\displaystyle \left|\frac{V_{ub}}{V_{cb}}\right| = 
\sqrt{\frac{m_u}{m_c}}\,,}\;\;\;\;\;\;\;\;\;
\Violet{\displaystyle \left|\frac{V_{td}}{V_{ts}}\right| =
\sqrt{\frac{m_d}{m_s}}},
$$
and the complex phase is given by
$$
\Violet{\cos \tilde \delta_{{\mathbb Z}_3} = -\frac{5}{8}}.
$$
The \Violet{Wolfenstein} parameter $\lambda$ has a dependence on the phase and
masses of light generations.

The status of experimental measurements for the elements of CKM matrix is
presented in Table \ref{Vckm} taken from \cite{Faccioli}. We see that the most
accurate determination is given for the element $V_{ud}$. Next, the element
$V_{cb}$ is measured with a low uncertainty, while the extraction of $V_{ub}$
involves model estimates, which can lead to underestimation of systematic
errors. This was recently demonstrated by M.Voloshin \cite{MBV}, who considered
the influence of factorization breaking in the calculation of hadronic matrix
elements for the four quark operators, that can result in the difference of
lifetimes for $D^0$ and $D_s$ mesons. These nonfactorizable effects can change
the form of end-point spectra in the decays of B mesons due to the $b\to u$
current, which is important in the theoretical description of corresponding
decays providing the extraction of $V_{ub}$. Finally, the ratio of
$V_{ts}/V_{td}$ is still constrained, but determined, because of difficulties
in the measuring of rapid $B_s$ oscillations.

\begin{table}[th]
\begin{center}
\caption{Present knowledge of the CKM matrix: Experimental determination
{[from Faccioli, 2000]}.}
\label{Vckm}
\renewcommand{\arraystretch}{1.2}
\footnotesize
\begin{tabular}{@{}clc@{}}
\hline\hline
$|V_{ij}|$ \scriptsize{etc.}   & \scriptsize{from}  &  \scriptsize{value}  \\
\hline
$G_{F}$   & \it{muon lifetime}  & $1.16639(1)\cdot 10^{-5}GeV^{-2}(\hbar
c)^{3}$ \\
$|V_{ud}|$  & \it{nuclear super-allowed decays}  &
$0.9740\pm0.0001_{exp}\pm0.0010_{th}$   \\
$|V_{ud}|$  & \it{neutron decay} & $0.9738\pm0.0016_{exp}\pm0.0004_{th}$
\\
$|V_{ud}|$  & \it{pion $\beta$ decay} & $0.9670\pm0.0160_{exp}\pm0.0008_{th}$
\\
\hline
$|V_{us}|$  & \it{$K_{e3}$ decays} & $0.2200\pm0.0017_{exp}\pm0.0018_{th}$  \\
$|V_{us}|$  & \it{hyperon semileptonic decays}  &  0.21 -- 0.24 \\
\hline
\hline
   $\begin{array}{c}
            |V_{cd}| \\ |V_{cs}|
         \end{array}$
&  \it{neutrino charm production}
&  $\begin{array}{c}
            0.225\pm0.012 \\ 1.04\pm0.16
         \end{array}$
\\

$|V_{cs}|$  & \it{$D_{e3}$ decays}  &  $1.02\pm0.05_{exp}\pm0.14_{th}$ \\
$|V_{cs}|$  & \it{hadronic $W$ decays} &  $0.99\pm0.02$        \\
\hline

$|V_{ub}|$     & $B \rightarrow \rho \ell \bar{\nu}$ & $(3.25 \pm 0.30_{exp}
\pm 0.55_{th}) 10^{-3}$ \\

$|V_{ub}/V_{cb}|$ &
      \renewcommand{\arraystretch}{1.0}
      \begin{tabular}{@{}l}
           \it {inclusive}  $B \rightarrow X_{u} \ell \bar{\nu}$ \\
           \it{\scriptsize (CLEO, ARGUS)}
      \end{tabular}
& $0.088 \pm 0.006_{exp} \pm 0.007_{th}$ \\

$|V_{ub}|^{2}$  & \it{inclusive} $b \rightarrow u \ell \bar{\nu}$ \scriptsize{\
(LEP)} & $(16.8 \pm 5.5_{exp} \pm 1.3_{th}) 10^{-6}$ \\

$|V_{cb}|$  & $B \rightarrow D^{\ast} \ell \bar{\nu}$  & $(42.8 \pm 3.3_{exp}
\pm 2.1_{th}) 10^{-3}$ \\
$|V_{cb}|$   & \it{inclusive} $b \rightarrow c \ell \bar{\nu}$  & $(41.2 \pm
0.7_{exp} \pm 1.5_{th}) 10^{-3}$ \\

\hline
\scriptsize{$|V_{tb}|^{2}/ \sum_{i}|V_{ti}|^{2}$} & \it{top quark decays} &
$0.93_{-0.23}^{+0.31}$ \\
\hline\hline

\multicolumn{3}{c}{\it effective FCNC processes} \\

\hline

$\begin{array}{c}
            |V_{td}V_{tb}| \\ |V_{ts}/V_{td}|
         \end{array}$
& \it{$B^{0}_{d}/\bar{B}^{0}_{d}$ and $B^{0}_{s}/\bar{B}^{0}_{s}$ oscillations}
&  $\begin{array}{c}
            (8.1\pm0.7_{exp}\pm0.6_{th}) \cdot 10^{-3} \\ > 4.6
         \end{array}$
\\

$|V_{ts}V_{tb}/V_{cb}|^{2}$ & \it{inclusive} $b \rightarrow s \gamma $  &
$0.94\pm0.11_{exp}\pm0.09_{th}$ \\
Im($V_{ij}$) & \it {CP-violation measurements:} &
\multicolumn{1}{l}{$|\epsilon_{K}| = (2.271 \pm 0.017) \cdot 10^{-3}$} \\
&& \multicolumn{1}{l}{$\epsilon'_{K}/\epsilon_{K} = (19.0 \pm 4.5) \cdot
10^{-4}$} \\
&& \multicolumn{1}{l}{$\sin 2 \beta = 0.48 _{-0.24}^{+0.22}$} \\

\hline\hline
\end{tabular}

\end{center}
\end{table}

In the heavy quark sector news come from the data acquisition at Belle and
BaBar experiments searching for the CP-violation in B decays \cite{BB}, which
we discuss below.

\subsection{Unitarity triangle}

The unitarity of mixing matrix in the SM provides us with useful conditions on
the values of matrix elements due to zero nondiagonal elements of
{$V_{CKM}\cdot V_{CKM}^\dagger=1$}. In B decays the corresponding condition can
be studied
$$
\Blue{ V_{ud} V_{ub}^* +  V_{cd} V_{cb}^* +  V_{td} V_{tb}^* =
0,}\hspace*{2cm}{\scriptstyle \times}\frac{1}{V_{cd} V_{cb}^*}.
$$
which, after the multiplication to the factor shown above, can be represented
as the equality for three vectors in the complex {\fbox{$\rho,\,\eta$} plane}
of {Wolfenstein parameters as shown in Fig. \ref{tri}.

\begin{figure}[th]
\begin{center}
\setlength{\unitlength}{1mm}
\begin{picture}(120,65)
\put(0,0){\epsfxsize=10cm \epsfbox{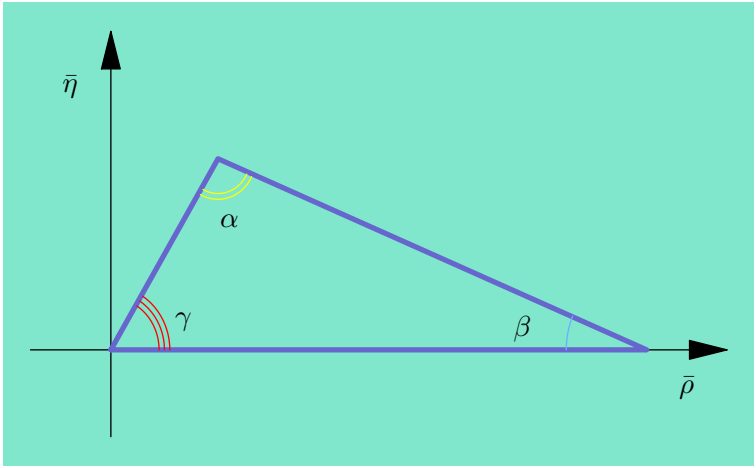}}
\put(90,10){$\bar\rho$}
\put(8,50){$\bar\eta$}
\put(68,17.5){$\beta$}
\put(23,19){$\gamma$}
\put(29,32){$\alpha$}
\end{picture}
\end{center}
\caption{The triangle.}
\label{tri}
\end{figure}

In terms of re-scaled \Violet{Wolfenstein} parameters, $\bar{\rho}=c\rho $,
$\bar{\eta}=c\eta $, $c=(1-\lambda ^{2}/2),$ the ratios $|V_{ub}|/|V_{cb}|$ and
$|V_{td}|/|V_{ts}|$ are 
\begin{equation}
{
\frac{|V_{ub}|}{|V_{cb}|}=\frac{\lambda }{c}\sqrt{\bar{\rho}^{2}+\bar{\eta},%
^{2}}\;\;\;\;\;\;\;\;\;\frac{|V_{td}|}{|V_{ts}|}\approx\frac{\lambda
}{c}\sqrt{(1-%
\bar{\rho})^{2}+\bar{\eta}^{2}} ,
}
\label{eq:vubcbtdts}
\end{equation}
which should be compared with the predictions based on the relations of mixing
parameters with the masses of fermions, i.e. on the regularities coming from
the principle of democracy
$$
\Violet{\displaystyle \left|\frac{V_{ub}}{V_{cb}}\right| = 
\sqrt{\frac{m_u}{m_c}}\,,}\;\;\;\;\;\;\;\;\;
\Violet{\displaystyle \left|\frac{V_{td}}{V_{ts}}\right| =
\sqrt{\frac{m_d}{m_s}}}.
$$

\begin{figure}[th]
\setlength{\unitlength}{0.7mm}
\hspace*{1cm}
\begin{picture}(170,90)
\put(0,0){\epsfxsize=11.9cm \epsfbox{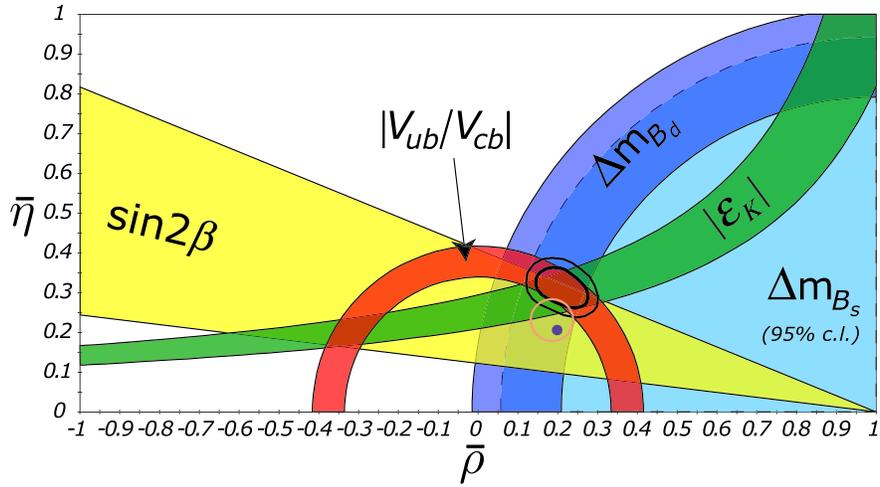}}
\put(104,33){\Melon{\circle{9}}}
\put(105,31){\Violet{\circle*{2}}}
\end{picture}
\caption{The position of triangle vertex: the experimental data compared with
the theoretical expectations at \Violet{$\cos \tilde \delta_{{\mathbb
Z}_3} = -\frac{5}{8}\Rightarrow $} \Violet{$\bullet$}, $\cos \tilde \delta =
-\frac{5}{8}\pm 0.045 \Rightarrow $ \Melon{$\bigcirc$}. \small Characteristic
values of quark masses at $\mu =m_Z$: $m_d\approx 4-4.5$ MeV, $m_u\approx
0.55\, m_d$, $m_s\approx 100-120$ MeV, $m_c\approx 0.65-0.67$ GeV, $m_b\approx
3-3.2$ GeV, $m_t\approx 181$ GeV.} 
\label{Fa}
\end{figure}
The position of triangle vertex in accordance with the current experimental
1$\sigma$-data on various parameters of CKM matrix \cite{Faccioli} is shown in
Fig.\ref{Fa}, where we also present the results of theoretical expectations
coming from the relations between the mass ratios of quarks and the mixings. We
see that allowing a light variation of CP-violating phase in the theoretical
model leads to a contour close to the present data on the unitarity triangle.

We can draw a conclusion that within the current accuracy we do not see any
contradiction of measurements with the mass-matrix hierarchy, moreover the
CP-violation phase is close to the value following from the {${\mathbb Z}_3$}
symmetry of vacuum.

\section{Neutral B mesons}
In this section we attribute the general characteristics of quark interactions
involving the CP-violating effects to the system of B mesons. Making common
remarks, we concentrate the attention to the neutral B mesons. First, we
describe how the forces breaking the CP parity manifest themselves in the
dynamics of flavor content in the neutral B mesons, i.e. in the static
mass-width parameters, as well as in the time evolution of flavored states.
Second, we show how the decay characteristics allow us to extract the
CP-violating effects in the quark interactions. Third, we attend some problems
in the theoretical interpretation of data as well as mention about the
consideration of CPT violation in the framework of dissipative dynamics.

\subsection{Eigen states}
We define the phase of combined CP-inversion, so that ${\sf CP}|B^0\rangle
\stackrel{\sf\scriptstyle def}{=} |\bar B^0\rangle $, where $B^0$ is flavored
state containing the $\bar b$ quark. Because of the mixing effects, the eigen
states of mass operator do not coincide with the flavored states. So, we denote
these states as
$$
|B_{\pm}\rangle = p|B^0\rangle\pm q|\bar B^0\rangle,
$$
while the mass matrix has the following general CPT-invariant form:
$$
{{\boldsymbol M}} - \frac{i}{2}{{\boldsymbol \Gamma}} =
\left(\begin{array}{cc}
M -\frac{i}{2}\Gamma & M_{12} - \frac{i}{2}\Gamma_{12}\\
M^*_{12} - \frac{i}{2}\Gamma^*_{12} & M -\frac{i}{2}\Gamma \end{array} \right),
$$
where ${\rm\boldsymbol M}$ is a dispersive part of mass matrix, and
${\boldsymbol \Gamma}$ is an absorbtive one. Solving the equation
$$
\left[{{\boldsymbol M}} - \frac{i}{2}{{\boldsymbol \Gamma}}\right]
|B_{\pm}\rangle = \lambda_{\pm}|B_{\pm}\rangle ,
$$
we get the positions of poles in the complex plane
$$
\lambda_\pm = \left(M-\frac{i}{2}\Gamma\right)\pm \frac{q}{p}
\left(M_{12}-\frac{i}{2}\Gamma_{12}\right),
$$
where
$$
\frac{q}{p} =
\sqrt{\frac{M^*_{12}-\frac{i}{2}\Gamma^*_{12}}{M_{12}-\frac{i}{2}\Gamma_{12}}}.
$$
In the SM we have got the nondiagonal mixing term
$$
M_{12} = - \Blue{\fbox{$(V^*_{td}V_{tb})^2$}}\; \PineGreen{G_F^2 M_W^2 M_B
f_B^2}\, \Violet{\eta_B(\alpha_s)}\,
B_B\, S(m_t/M_W),
$$
where the framed factor represents the dynamics of charged current mixing,
$\Violet{\eta_B(\alpha_s)}$ includes the QCD corrections to the quark level
diagrams, $B_B$ gives the deviation for the hadronic matrix element of
four-quark operator from the factrorized expression in terms of hadronic matrix
elements for two-quark operators (the leptonic constant $f_B$). The function
$S(m_t/M_W)$ is known and calculated at the quark level. The absorptive part of
mixing is also known, and it is suppressed as
\begin{equation}
\left|\frac{\Gamma_{12}}{M_{12}}\right|\sim {\cal O}(m_b^2/m_t^2) \ll 1.
\label{gm}
\end{equation}
Alternative notations analogous to the K meson physics are also usually
explored
$$
|B_{\pm}\rangle = \frac{(1+\epsilon)|B^0\rangle \pm(1-\epsilon)|\bar
B^0\rangle}{\sqrt{2(1+|\epsilon|^2)}},
$$
with
$$
\frac{1-\epsilon}{1+\epsilon} = \frac{q}{p}.
$$
Due to the nonzero absorptive part, the absolute value of ratio representing
the fractions of flavored states in the massive ones deviates from unity, so
that to the subleading order of small parameter (see (\ref{gm})) we get 
$$
\left|\frac{q}{p}\right|^2 =1 + \left|\frac{\Gamma_{12}}{M_{12}}\right|
\sin[arg(M_{12})-arg(\Gamma_{12})]+\ldots
$$
We can point to the conditions, when the CP-violation effects are absent in the
SM with three generations,
$$
\left.\begin{array}{r}
\displaystyle \left|\frac{q}{p}\right| =1 \\[5mm]
{\mathfrak Re} [\epsilon] = 0
\end{array}
\right\} \Longrightarrow \mbox{no {CP} violation}\;\; \Leftrightarrow  \;\;
CP |B_{\pm}\rangle =\pm\,|B_{\pm}\rangle .
$$
In that case the massive states are orthogonal to each other and have definite
CP parities.
\subsection{Time evolution}
After the production of flavored states at the moment $t=0$: $|B^0(0)\rangle =
|B^0\rangle$ and $|\bar B^0(0)\rangle = |\bar B^0\rangle $, the evolution of
flavor contents takes place, so that it can be expressed in terms of
eigen-values of mass operator, and we obtain
\begin{eqnarray}
|B^0(t)\rangle &=& g_+(t)|B^0\rangle +\frac{q}{p}\, g_-(t) |\bar B^0\rangle,
\nonumber\\[2mm]
|\bar B^0(t)\rangle &=& g_+(t)|\bar B^0\rangle +\frac{p}{q}\, g_-(t)
|B^0\rangle, \nonumber
\end{eqnarray}
where
$$
g_\pm(t) = \frac{1}{2} \left(e^{-i\lambda_+ t} \pm e^{-i\lambda_- t}\right),
$$
and
$$
|g_\pm(t)|^2 = \frac{e^{-\Gamma t}}{2}\left[ \mbox{cosh}
\left(\frac{\Delta\Gamma}{2}\, t\right)\pm \cos(\Delta m\, t)\right],
$$
with
$$
\Delta \Gamma = |\Gamma_+-\Gamma_-|,\hspace*{1cm}
\Delta m = |M_+-M_-|.
$$
We see that \underline{almost coherent oscillations of $B^0\leftrightarrow \bar
B^0$ untill decay} take place, and the deviation is given by the non-unit
absolute value of factor $|q/p|$.

If both decays of $B^0-\bar B^0$ system are tagged by their $b$-flavor
contents, then putting $\Delta \Gamma t\to 0$ we have got for the tagged events
$$
|g_\pm(t)|^2\;  \frac{2}{e^{-\Gamma t}} = \left[ 1 \pm \cos(\Delta m\,
t)\right],
$$
which is measured experimentally (see Fig.\ref{BaB}).
\begin{figure}[th]
\centerline{\epsfxsize=8cm\epsfbox{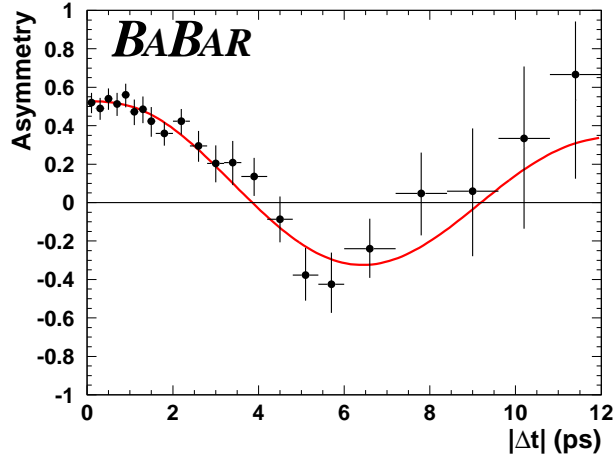}}
\caption{The {BaBar} data on the oscillation of tagged events. The asymmetry is
defined as $(|g_+(t)|^2 - |g_-(t)|^2)/(|g_+(t)|^2 + |g_-(t)|^2)$, while the
deviation of amplitude from the unity is due to so-called dilution factor of
mistagging.}
\label{BaB}
\end{figure}

\subsection{Decays \& CP-vilation}
The main feature of dynamics under consideration is that in order to observe
the \Blue{CP violation} we have to involve the \Violet{\fbox{interference}}.

One usually isolates three classes of processes under interest:
\begin{itemize}
\item
\Red{{\bf Indirect CP Violation}}: $ \displaystyle{\left|\frac{q}{p}\right|
\neq 1}$ is enough to measure the CP-violation.
\item
\Red{{\bf Direct CP Violation I}}: Asymmetry in flavored channel decay
amplitudes with \\ \hspace*{4.4cm}$|A(f)| \neq |A^{CP}(f^{CP})|$ or different
weak phases 
in two decays.
\item
\Red{{\bf Direct CP Violation II}:} {{\sf Decays to CP eigenstates}}:
{\begin{itemize}\item \fbox{Interference} of oscillations with decay
amplitudes even at {$$\left|\frac{q}{p}\right|=\tilde q = 1\;\;\;\;
\mbox{and}\;\;\;\; |A(f)| = |A^{CP}(f^{CP})|.$$} \end{itemize}}
\end{itemize}
These classes correspond to the following decay processes:

1. Wrong-flavor decays, for example, the wrong lepton-sign widths give
$$
a_{SL}^{CP} = \frac{\Gamma(\bar B\to l^+ X)-\Gamma(B\to l^- X)}{\Gamma(\bar
B\to l^+ X)+\Gamma(B\to l^- X)} = \frac{1-\tilde q^4}{1+\tilde q^4}. 
$$

2. The asymmetry in decays to flavored channels (charged $B_q$ mesons, too)
occurs, when the weak CP-violating phases of amplitudes interfere with the
terms of CP-even phases of strong interaction.

3. The time-dependent asymmetry in decays of neutral B mesons to CP eigenstates
represents the normalized difference of events in decays of states with
definite $b$-flavor at zero time, which is determined by tagging the flavor in
the decay of associated neutral B meson. So, it has the form
\begin{equation}
a^{CP}(t) = {\mathfrak Im}\left[\frac{q}{p}\, \frac{A^{CP}}{A} \right]
\sin[\Delta m\, t],\;\;\;\;\;\; A = A(B\to f).
\label{asy}
\end{equation}
Following the items 2 and 3, one needs quite a definite understanding of decay
amplitude structure by isolating various dynamical factors such as the phases
of weak and strong interaction terms.

The CP-odd \Violet{phases} of amplitudes are \PineGreen{classified} by the
\Red{effective weak lagrangian of heavy quark decays}.

\begin{figure}[th]
\setlength{\unitlength}{0.7mm}
\begin{center}
\begin{picture}(160,86)
\put(0,17){\epsfxsize=6.3cm \epsfbox{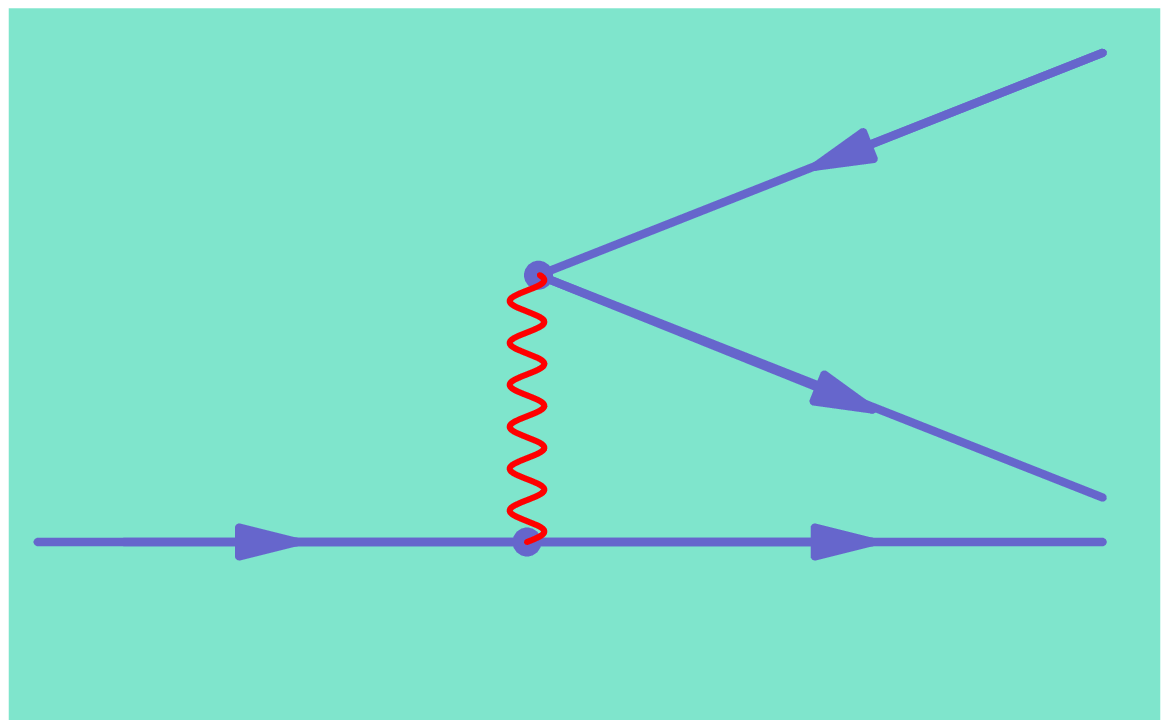}}
\put(90,0){\epsfxsize=4.2cm \epsfbox{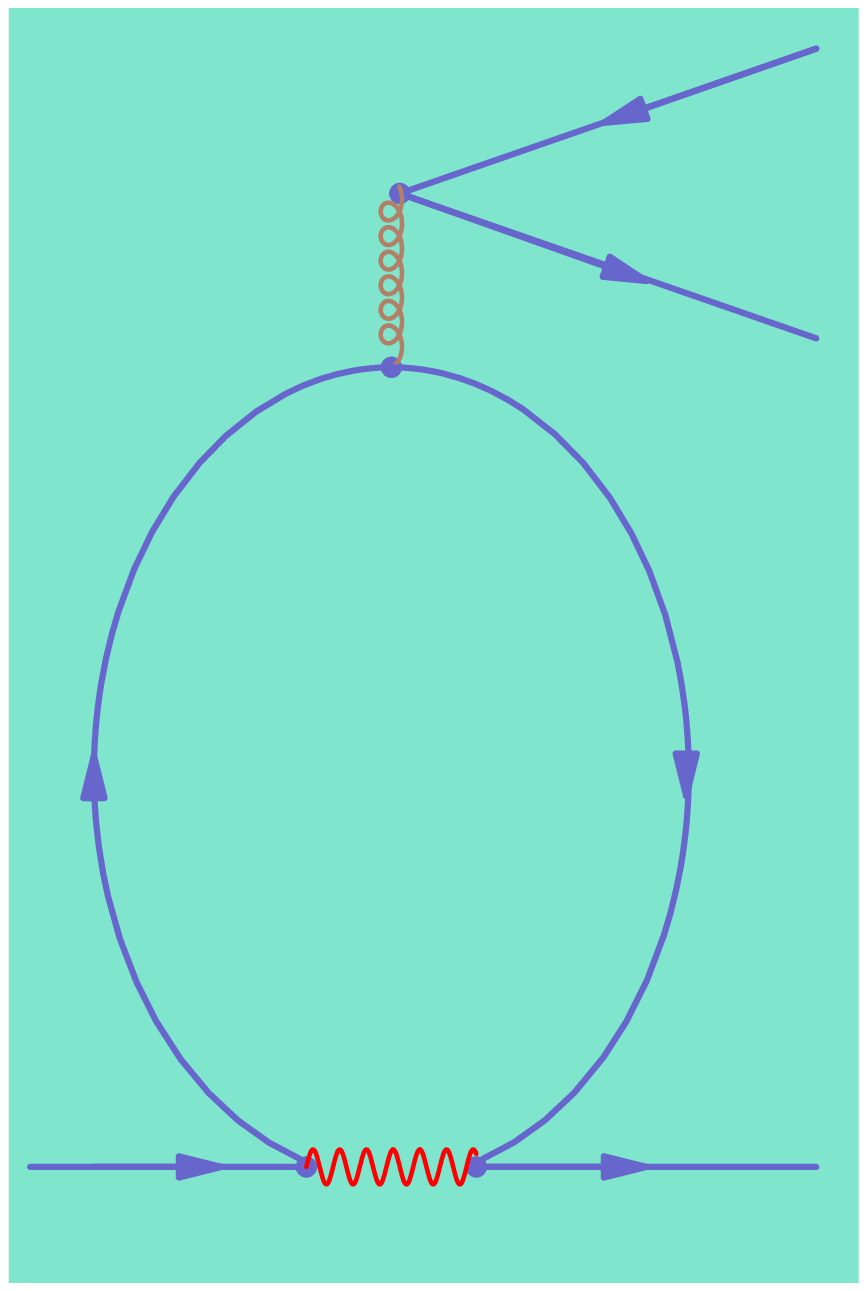}}
\put(10,25){$\boldsymbol b$}
\put(70,25){$\boldsymbol c$}
\put(70,43){$\boldsymbol s$}
\put(70,67){$\boldsymbol {\bar c}$}
\put(32,40){$\boldsymbol W$}
\put(95,2){$\boldsymbol b$}
\put(115,2){$\boldsymbol W$}
\put(135,2){$\boldsymbol s$}
\put(112,70){$\boldsymbol g$}
\put(135,65){$\boldsymbol c$}
\put(135,85){$\boldsymbol {\bar c}$}
\put(123,35){$\boldsymbol {u,c,t}$}
\end{picture}
\end{center}
\caption{The diagrams representing various weak phase structures in the
effective four-fermion weak lagrangian of quarks, i.e. the tree diagram and the
penguin. The gluon corrections, which do not change these phases, are not
shown, but they are known to two-loop order in QCD coupling $\alpha_s$
\cite{RMPhys}.}
\label{ccs}
\end{figure}

The corresponding diagrams in the decays of $b\to c\bar c s$ as they contribute
to the effective weak lagrangian of quarks are shown in Fig.\ref{ccs}. Some
other gluon corrections are also calculable (see review in \cite{RMPhys}).
Various weak CP-violating terms are arranged in powers of small parameter, the
sine of Cabibbo angle $\lambda$, and presented in Table \ref{prdt}.

\begin{table}[th]
\caption{The quark processes in decays of neutral B mesons and their arrangment
in $\lambda$ with the indication of appropriate angle in the unitarity
triangle. The symbol $^\star$ means that the angle would be extracted if the
competitive term of correction would be small.}
\label{prdt}
\begin{tabular}{cp{4.5cm}p{4.5cm}p{2cm}c}
\hline
Current & Leading term & Correction & $B_d$ decay mode & weak angle \\[2mm]
\hline
$b\to c\bar c s$ & \mbox{$V_{cb} V_{cs}^* = A \lambda^2$} \mbox{\Red{tree} +
\Blue{penguin}} (c-t) & $V_{ub} V_{us}^* = A \lambda^4
\scriptstyle(\rho-i\eta)$ \Blue{penguin} (u-t) & $J/\Psi\, K_S$ & $\beta$
\\[2mm]
\hline
$b\to s\bar s s$ & \mbox{$V_{cb} V_{cs}^* = A \lambda^2$} \mbox{\Blue{penguin
only}}
(c-t) & $V_{ub} V_{us}^* = A \lambda^4 \scriptstyle(\rho-i\eta)$ \Blue{penguin}
(u-t) & $\phi\, K_S$ & $\beta$ \\[2mm]
\hline
$b\to c\bar c d$ & \mbox{$V_{cb} V_{cd}^* = -A \lambda^3$} \mbox{\Red{tree} +
\Blue{penguin}} (c-u) & $V_{tb} V_{ts}^* = A \lambda^3
\scriptstyle(1-\rho+i\eta)$
\Blue{penguin} (t-u) & $D^+\, D^-$ & $^\star\beta$ \\[2mm]
\hline
$b\to c\bar u d$ & \mbox{$V_{cb} V_{ud}^* = A \lambda^2$}
\mbox{\Red{tree~~~~~~~}} & 0 & $D^0\, \pi^0 (\rho^0)$
\mbox{~$^|\hspace*{-2mm}\to $
CP eigen}
\mbox{~~~~~~~~state} & $\beta$ \\[2mm]
\hline
$b\to u\bar u d$ & \mbox{$V_{ub} V_{ud}^* = A \lambda^3 \scriptstyle
(\rho-i\eta)$} \mbox{\Red{tree} + \Blue{penguin}} (u-c) &
$V_{tb} V_{ts}^* = A \lambda^3 \scriptstyle(1-\rho+i\eta)$
\Blue{penguin} (t-c) & $\pi\pi;\; \rho\pi$ & $^\star\alpha$ \\[2mm]
\hline
\end{tabular}
\end{table}

As an example we point to the so-called \Blue{Golden plated mode} {$J/\Psi
+ K_S$}. In this case the amplitudes are determined theoretically with
extremely low uncertainty (below 1\%), since the corrections to the amplitudes
are suppressed by the sine of Cabibbo angle squared.

Following the equation for the asymmetry, we can determine the quantities
entering (\ref{asy}) and their weak CP-odd phases:
$$
\begin{array}{ccc}
\displaystyle \frac{q}{p} & \Longrightarrow  &
\displaystyle\frac{V^*_{tb}V_{td}}{V_{tb}V^*_{td}}, \\[4mm]
\displaystyle\frac{A_B^{CP}}{A_B}  & \Longrightarrow  &
\displaystyle\frac{V^*_{cs}V_{cb}}{V_{cs}V^*_{cb}},
\\[4mm]
K_S & \Longrightarrow  & \displaystyle\frac{V^*_{cd}V_{cd}}{V_{cd}V^*_{cd}},
\end{array}
$$
so that we get
$$
a^{CP}_{J/\Psi K_S}(t) = \sin[2\beta]\, \sin[\Delta m\, t].
$$
This quantaty is observed in various experiments, while the Belle and BaBar are
tending to improve the measurement of angle $\beta$. The summary of current
data on $\sin 2\beta$ is presented in Fig.\ref{2b}.
\begin{figure}[th]
\centerline{\epsfxsize=6.5cm \epsfbox{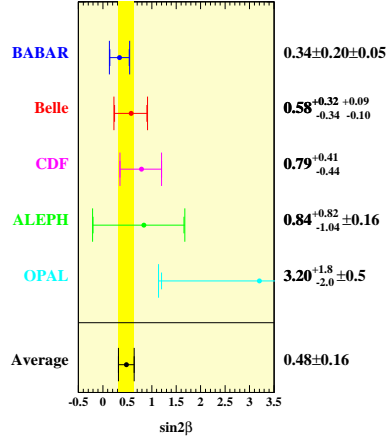}}

\vspace*{-4mm}
\caption{The data on $\sin 2\beta$ and the world-average value.}
\label{2b}
\end{figure}

\subsection{Some problems}
In contrast to the golden plated mode the other decay channels for B mesons
cannot be analyzed with a high accuracy of theoretical description. In this
section, first, we point to some difficulties in the extraction of weak angles
$\alpha,\; \gamma$ from the data, and second, we mention a general possibility
of violation in the invariance of interactions under the complete CPT
conjugation.

\vspace*{3mm}
\noindent
{\fbox{Mode $B\to \pi\pi$}}

In this decay the ratio of amplitudes for the CP conjugated initial states with
definite flavor is not well determined theoretically because of significant
contribution by the penguin with the different weak phase. The magnitude of
this penguin depends on the strong dynamics of light quarks bound inside the
mesons, which is described with large numerical uncertainties or even
model-dependent.

\noindent
The problem is connected to the evaluation of \underline{hadronic matrix
elements for the quark currents}:
\begin{itemize}
\item
{the factorization} \& \Blue{nonfactorizable effects} in the evaluation of
four quark operators,
\item
unknown relative strong phases between the contributions
possessing different weak CP-odd phases,
\item
{the isotopic symmetry} could help, but $\pi^0\pi^0$ channel is difficult to
measure,
\item
one has to use \underline{dynamical, theoretical predictions}, that leads to
\Blue{uncertainties, model-depen\-dent formfactors}.
\end{itemize}
The analysis of above problems was carefully performed in \cite{Beneke}. A
characteristic picture following from such the calculations is shown in
Figs.\ref{b1} and \ref{b2}. The restrictions presented in Fig.\ref{b1} are
separated in two classes. The first is the shaded region as descrined in
Section 2. The second class is the analysis of $\pi\pi$ and $K\pi$ modes in
\cite{Beneke}, that is shown as 1, 2 and 3 sigma regions as well as the dots
giving both the factorization-based result and the model dependent evaluation
of nonfactorizable effects. Fig.\ref{b2} shows the experimental bounds on the
various ratios of decay widths under consideration (horizontal bands) in
comparison with the factorization result (dashed curve) as well as the
reasonable variation of form factors (shaded regions) versus the weak angle
$\gamma$ \cite{Beneke}.
\begin{figure}[th]
\centerline{\epsfxsize=8cm \epsfbox{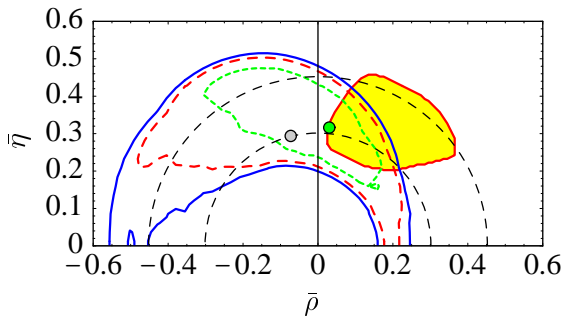}}
\caption{The analysis of restrictions following from $B\to \pi\pi$ and $B\to
K\pi$ decays \cite{Beneke}. The explanations are given in the text.}
\label{b1}
\end{figure}

\noindent
{\fbox{Mode $B\to \rho(\pi\pi)\pi$}}

\vspace*{4mm}
\noindent
For this mode in comparison with the $\pi\pi$ channel, the additional problem
is the presence of strong resonances in two-pion states. This involves a large
uncertainty because of increase of unknown parameters such as the strong phases
and absolute values of various amplitudes. However, one can explore the
advantage of \underline{Dalitz plot analysis}:
\begin{itemize}
\item
\Red{a hope} to \Blue{extract all of phases and amplitudes} \cite{SnyderQuinn},
\item
to recognize uncertainties of reconstruction: 
\begin{itemize}
\item
many resonances,
\item
{nonresonant} contribution.
\end{itemize}
\end{itemize}

\begin{figure}[ph]
\centerline{\epsfxsize=11cm \epsfbox{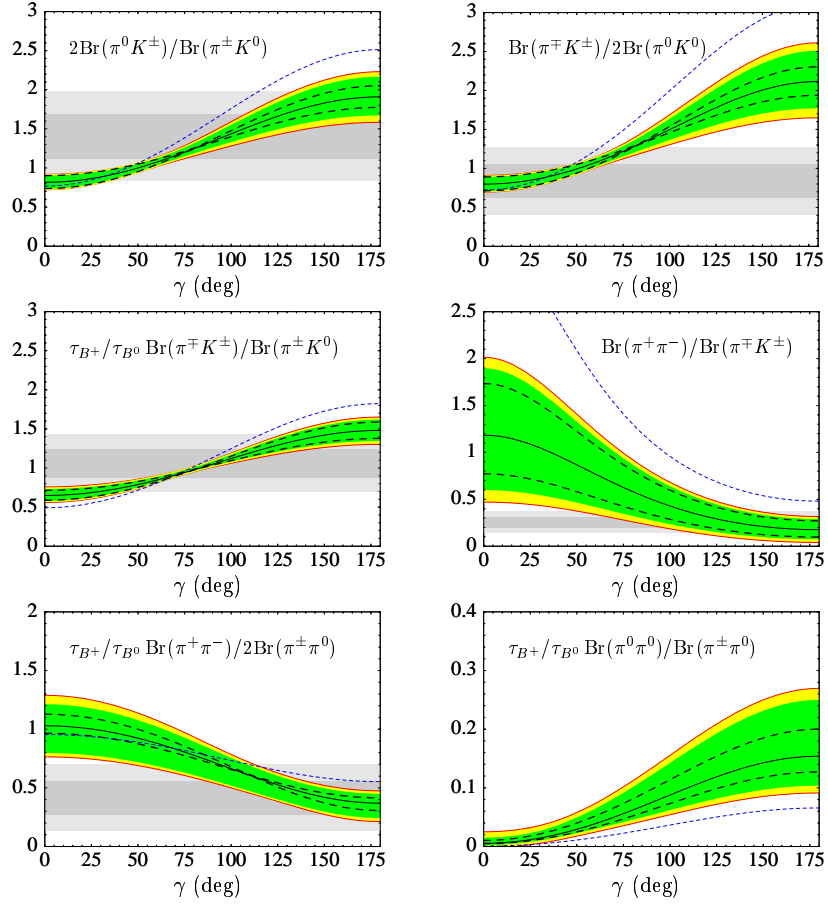}}
\caption{The comparison of experimental bounds with the theoretical predictions
on $B\to \pi\pi$ and $B\to K\pi$ decays \cite{Beneke}. The explanations are
given in the text.}
\label{b2}
\end{figure}
\begin{figure}[ph]
\epsfxsize=7.5cm \raisebox{-1mm}{\epsfbox{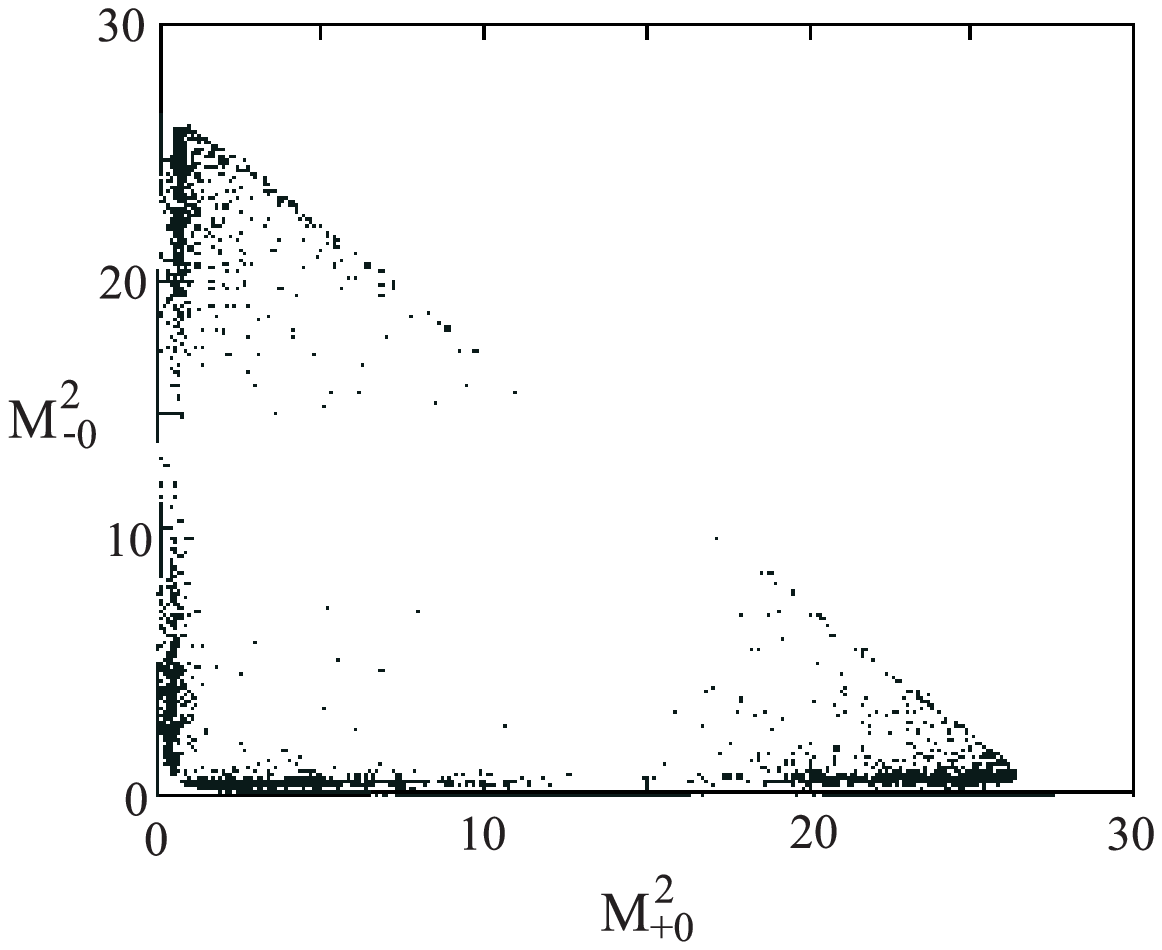}}
\hspace*{1cm}\epsfxsize=6cm \epsfbox{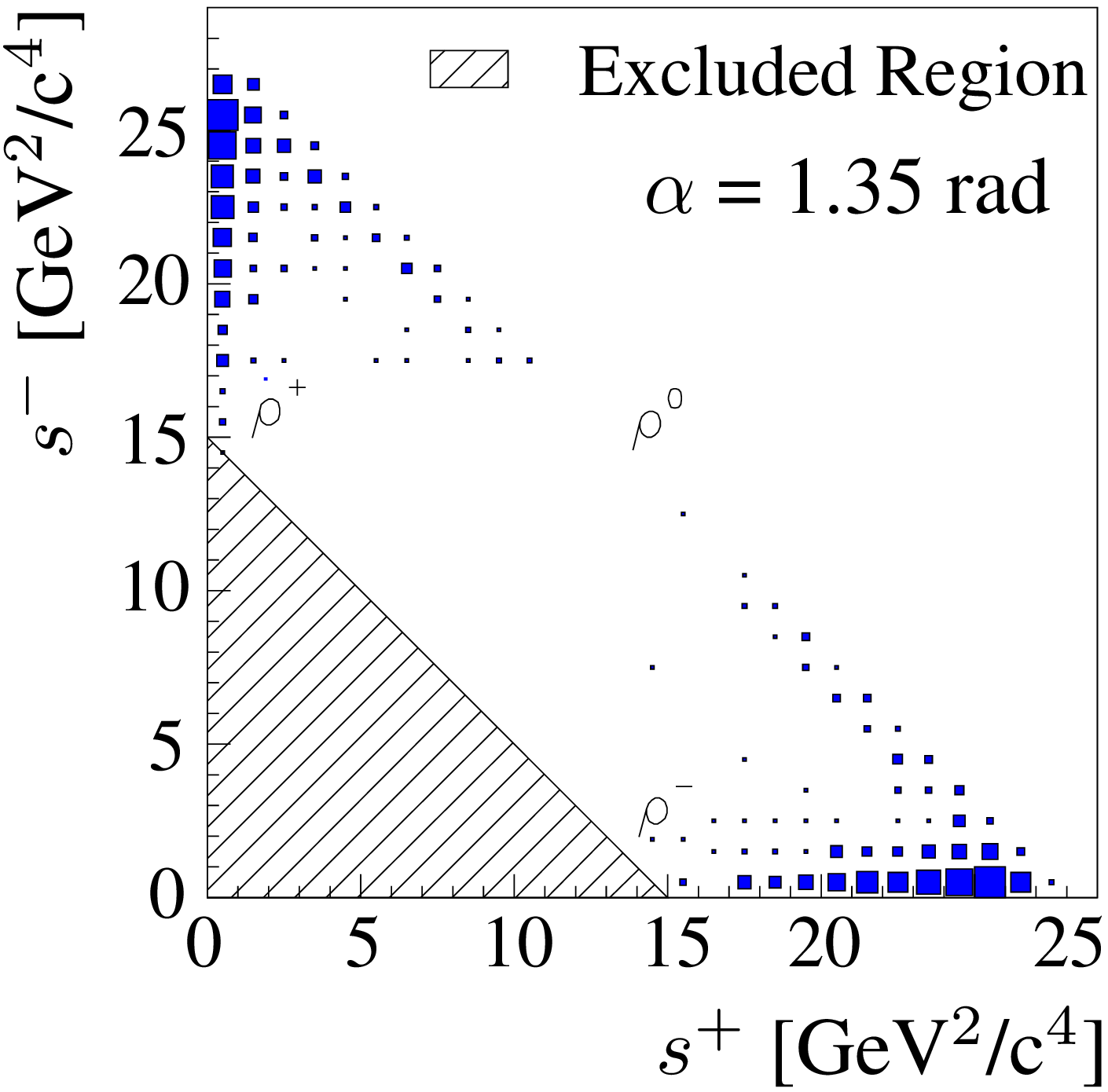}
\caption{The Dalitz plots in $B\to 3\pi$ decays simulated in \cite{SnyderQuinn}
(left figure) \& at BTeV, LHCb (right figure).}
\label{sny}
\end{figure}

\noindent
The results of simulations performed in the original paper by Snyder \& Quinn
\cite{SnyderQuinn} as well as the expected picture at BTeV and LHCb are shown
in Fig.\ref{sny}. An error of such the analysis in the determination of angle
$\gamma$ is expected on the level of less than $10^\circ$.

\vspace*{4mm}
\noindent
\Blue{\fbox{CPT violation}}

\vspace*{4mm}
Another problem is that the CP and T conjugations could be nonequivalent if the
complete CPT inversion is violated in the interactions. This could happen, for
example, if we deal with \underline{incomplete dynamics}:
\begin{itemize}
\item
\underline{extra dimensions} could results in that \Blue{4D conversions} do not
lead to equvalent representations of \underline{extended `Poincare' group},
that implies at low energies with the effective four dimensional interactions
the extended dynamics would cause the CPT-violation,
\item
a general scheme with a \underline{dissipative Hamiltonian} resulting in
description in terms of quantum dynamical semigroups was considered in
\cite{BenattiFloreaniniRomano}.
\end{itemize}
This dissipative dynamics involves {\bf 6} additional parameters of correction
$L$ to the mass matrix of neutral B mesons under the conditions of
\underline{conserving the positivity and entropy growth}, so that
$$
{{\boldsymbol M}} - \frac{i}{2}{{\boldsymbol \Gamma}}
\;\;\; \Longrightarrow \;\;\; {{\boldsymbol M}} - \frac{i}{2}{{\boldsymbol
\Gamma}} + {L}.
$$
In that case the asymmetries of T-inverted amplitudes $A$ are different from
the CP-conjugated ones, while the asymmetry with CPT conversion is not equal to
zero, if the diagonal elements of $L$ do not coincide with each other:
$$
A_T \neq A_{CP},\;\;\;A_{CPT}\neq 0\;\;\; \Longleftarrow \;\;\; L_{11} \neq
L_{22}.
$$
One expects for rather strict limits on the {{\bf\small CPT}}-violating
parameters from current experiments \cite{BenattiFloreaniniRomano}.

\section{Conclusion}
To summarize we draw the following conclusions:
\begin{itemize}
\item
Mass matrices of fermions govern the charged current mixings:
\begin{itemize}
\item
we observe the {\bf charged fermion mass hierarchy}, and
\item
the {\bf mixing hierarchy},
\end{itemize}
which can be presented as a common property of Yukawa interactions in the form
of democratic (almost equal) couplings to the scalar vacuum field. 
\item
Small perturbations of \Red{generation democracy } result in the following:
\begin{itemize}
\item
the {\bf mixing angles} are related with the {\bf mass ratios},
\item
the ${\mathbb Z}_3$ symmetry of vacuum leads to a definite {CP-phase} of mixing
matrix, 
\item
the current knowledge of CKM matrix is in agreement with the mass relations and
very close to the ${\mathbb Z}_3$ symmetry of vacuum.
\end{itemize}
\item
\underline{Golden plated mode brings the angle $\beta$ of unitarity triangle}.
\item
Strong {theoretical \& experimental} efforts are challenged to extract other
angles. 
\end{itemize}
Some other aspects of CP-violation in B decays are discussed in reviews
\cite{CPrev}.

This work is in part supported by the Russian Foundation for Basic Research,
grants 01-02-99315, 01-02-16585 and 00-15-96645.

\end{document}